\documentclass[reprint,aps,prd,superscriptaddress,twocolumn,longbibliography,notitlepage,nofootinbib,10pt]{revtex4-2}
\usepackage{color}
\definecolor{color1}{rgb}{0,0,0.7}
\definecolor{color2}{rgb}{0.85,0,0}

\usepackage[breaklinks=true,colorlinks,citecolor=red,linkcolor=color2,urlcolor=color1]{hyperref}
\usepackage{graphicx}
\usepackage{amsmath}
\usepackage{amssymb}
\usepackage{amsfonts}
\usepackage{amsthm}
\usepackage{bbm}
\usepackage{soul}
\usepackage{textcomp}
\usepackage{bm}
\usepackage{dsfont}
\usepackage{relsize}
\usepackage{braket}
\usepackage{enumitem}
\usepackage{cleveref}
\usepackage{float}
\usepackage[makeroom]{cancel}

\usepackage{pifont}

\usepackage{titlesec}
\titlespacing*{\section}{0pt}{10pt plus 4pt minus 5pt}{7pt plus 3pt minus 5pt}
\titlespacing*{\subsection}{0pt}{5pt plus 3pt minus 1pt}{4pt plus 2pt minus 4pt}
\titleformat{\section}{\centering\bfseries}{\thesection.}{.5em}{}

\newcommand{\eref}[1]{\textcolor{color2}{\hyperref[#1]{eq.$\,$(\ref{#1})}}}
\newcommand{\Eref}[1]{\textcolor{color2}{\hyperref[#1]{Eq.$\,$(\ref{#1})}}}
\newcommand{\fref}[1]{\textcolor{color2}{\hyperref[#1]{Fig.$\,$\bfseries\ref{#1}}}}

\newcommand{\sfref}[2]{\textcolor{color2}{\hyperref[#1]{Fig.$\,$\bfseries\ref{#1}(#2)}}}
\newcommand{\tref}[1]{\textcolor{color2}{\hyperref[#1]{Table~\bfseries\ref{#1}}}}

\newcommand{\aref}[1]{\textcolor{color2}{\hyperref[#1]{App.$\,$\ref{#1}}}}
\newcommand{\sref}[1]{\textcolor{color2}{\hyperref[#1]{Sec.$\,$\ref{#1}}}}

\def\eps{\varepsilon}

\DeclareMathOperator*{\argmin}{arg\,min}

\newcommand{\nocontentsline}[3]{}

\begin{document}
	
\title{Rapid optimal work extraction from a quantum-dot information engine}

\author{Kushagra Aggarwal}
\altaffiliation{These authors have contributed equally to this work.}
\affiliation{Department of Engineering Science, University of Oxford, Parks Road, Oxford OX1 3PJ, United Kingdom}

\author{Alberto Rolandi}
\altaffiliation{These authors have contributed equally to this work.}
\affiliation{Département de Physique Appliquée, Université de Genève, 1211 Genève, Switzerland}
\affiliation{Atominstitut, TU Wien, 1020 Vienna, Austria}

\author{Yikai Yang}
\affiliation{Department of Engineering Science, University of Oxford, Parks Road, Oxford OX1 3PJ, United Kingdom}

\author{Joseph Hickie}
\affiliation{Department of Materials, University of Oxford, Parks Road, Oxford OX1 3PH, United Kingdom}

\author{Daniel Jirovec}
\affiliation{Institute of Science and Technology Austria, Am Campus 1, 3400 Klosterneuburg, Austria}

\author{Andrea Ballabio}
\affiliation{L-NESS, Physics Department, Politecnico di Milano, via Anzani 42, 22100, Como, Italy}

\author{Daniel Chrastina}
\affiliation{L-NESS, Physics Department, Politecnico di Milano, via Anzani 42, 22100, Como, Italy}

\author{Giovanni Isella}
\affiliation{L-NESS, Physics Department, Politecnico di Milano, via Anzani 42, 22100, Como, Italy}

\author{Mark T. Mitchison}
\affiliation{School of Physics, Trinity College Dublin, College Green, Dublin 2, D02 K8N4, Ireland}
\affiliation{Department of Physics, King’s College London, Strand, London, WC2R 2LS, United Kingdom}

\author{Martí Perarnau-Llobet}
\email{marti.perarnau@uab.cat}
\affiliation{F\'isica Te\`orica: Informaci\'o i Fen\`omens Qu\`antics, Department de F\'isica, Universitat Aut\`onoma de Barcelona, 08193 Bellaterra (Barcelona), Spain}
\affiliation{Département de Physique Appliquée, Université de Genève, 1211 Genève, Switzerland}

\author{Natalia Ares}
\email{natalia.ares@eng.ox.ac.uk}
\affiliation{Department of Engineering Science, University of Oxford, Parks Road, Oxford OX1 3PJ, United Kingdom}

	\begin{abstract}    
    The conversion of thermal energy into work is usually more efficient in the slow-driving regime, where the power output is vanishingly small. Efficient work extraction for fast driving protocols remains an outstanding challenge at the nanoscale, where fluctuations play a significant role. In this Letter, we use a quantum-dot Szilard engine to extract work from thermal fluctuations with maximum efficiency over two decades of driving speed. We design and implement a family of optimised protocols ranging from the slow- to the fast-driving regime, and measure the engine's efficiency as well as the mean and variance of its power output in each case. These optimised protocols exhibit significant improvements in power and efficiency compared to the naive approach. Our results also show that, when optimising for efficiency, boosting the power output of a Szilard engine inevitably comes at the cost of increased power fluctuations.
	\end{abstract}
	
	\maketitle
	
Converting thermal energy into work is the central problem of thermodynamics. Efficient work extraction requires quasi-static operations with vanishing average power output.~\cite{Shiraishi2016}. This trade-off between efficiency and power also extends to the fluctuations, which diverge to achieve finite power at Carnot efficiency~\cite{Campisi2016,Pietzonka2018}. This motivates the need for efficient protocols under finite-time, far-from-equilibrium conditions. 

Information engines, such as the paradigmatic Szilard engine~\cite{szilard1964decrease}, provide an ideal setting to address this problem. Information obtained through measurement can, in principle, fully convert thermal energy to work~\cite{Landauer1991,Parrondo2015,duBuisson2024}. In an information engine, therefore, any loss of efficiency results from the non-equilibrium nature of the driving protocol. By contrast, in conventional heat-driven engines, efficiency is limited both by non-equilibrium effects and by the need to dump some energy into a cold bath to ensure consistency with the second law of thermodynamics. Inferring thermal engine efficiency requires independent measurements of heat and work, a persistent challenge at the nanoscale despite recent progress~\cite{Josefsson2018, Karimi2020, Majidi2022,Majidi2024}.

While information-to-work conversion has a long history, dating back to seminal ideas of Maxwell, Szilard and Landauer~\cite{Maxwell11,landauer1961irreversibility,szilard1964decrease}, its experimental implementation is more recent and has been driven by developments in the fields of stochastic and quantum thermodynamics~\cite{Ciliberto2017,Myers2022}. Pioneering experiments have realised a Szilard engine~\cite{Toyabe2010} and the erasure of information close to the Landauer limit~\cite{Brut2012} on single colloidal particles. Subsequent experiments explored the link between information and thermodynamics in Brownian colloidal particles~\cite{Jun2014High,Paneru2018,Admon2018} and other platforms such as quantum dots~\cite{koski_experimental_2014, Barker2022, Scandi2022}, ultracold atoms~\cite{Kumar2018}, quantum memories~\cite{Camati2016} and superconducting circuits~\cite{Cottet2017,Masuyama2018}. However, while optimal extraction of work from information has been demonstrated in the slow-driving regime~\cite{Scandi2022}, the more challenging case of arbitrary driving speeds is yet to be addressed. 

Here we address this challenge by experimentally implementing optimal work extraction in a quantum-dot Szilard engine, realizing optimal protocols over two orders of magnitude in driving speed. The optimal protocols are found by generalizing the results of Ref.~\cite{Esposito2010} to our system. These interpolate between the two-jump protocols for fast driving~\cite{Blaber2021Steps,Rolandi2023Optimal} and the geodesic protocols of Ref.~\cite{scandiThermodynamicLengthOpen2019} for slow driving. We also characterise work fluctuations, which play a dominant role at these scales~\cite{Jarzynski2011}. We observe that, whereas in the high efficiency regime work fluctuations disappear due to the fluctuation-dissipation relation, higher power comes inevitably with higher fluctuations. We demonstrate an information engine in planar germanium, a scalable platform for classical and quantum information processing~\cite{Scappucci2020}.

\begin{figure}
    \centering
    \includegraphics{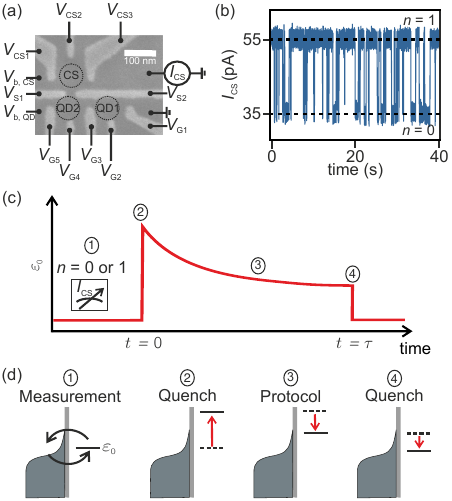}
    \caption{(a) Scanning electron microscope image of the device. QD1 and QD2 are controlled using voltages $V_\text{G1-G5}$. QD1 encodes the information bit. The charge sensor quantum dot (CS) is defined using voltages $V_\text{CS1-CS3}$. (b) A time trace of the current $I_\text{CS}$ through the charge sensor showing the stochastic tunnelling of a particle in the information quantum dot. (c) In the first step, the quantum dot is initialised at $50\%-50\%$ and pulsed to a known state. (d) It is reset to the initial state and in process gaining energy, realising a Szilard engine.}
    \vspace*{-5pt}
    \label{fig:fig1}
\end{figure}
	
\emph{Experiment.} Our device is shown in \sfref{fig:fig1}{a}. It consists of a quantum dot defined in a germanium quantum well. An information bit is encoded in the occupancy of the right dot in the bottom array (QD1). QD1 is defined by voltages $V_\text{G1}$ and $V_\text{G3}$. The quantum dot's electrochemical potential $\eps$ is controlled using voltage $V_\text{G2}$. The left dot (QD2) in the array is tuned in Coulomb blockage, limiting the tunnelling of QD1 to the right reservoir. Another quantum dot in the top array (CS) electrostatically coupled to QD1, defined using gates $V_\text{CS1-CS3}$, serves as a probe for QD1's occupancy $n$. The occupancy $n$ is monitored by measuring the current $I_\text{CS}$ as shown in \sfref{fig:fig1}{b}.  
	
The experiment is performed in a regime where $n \in \{0,1\}$, i.e. when QD1 has an effective occupation of 0 or 1. The tunnelling rates are $\gamma_\text{in} = \Gamma_\text{in}f(\eps)$ and $\gamma_\text{out} = \Gamma_\text{out}(1-f(\eps))$, with $\Gamma_\text{in} = 7.0\ \text{Hz}$ and $\Gamma_\text{out} = 3.5\ \text{Hz}$, where $f(x) = (1+e^{\beta x})^{-1}$ is the Fermi function and $\beta = (k_B T)^{-1}$ is the inverse of the system's temperature ($T = 180~\text{mK}$). We note that $\Gamma_\text{in} \approx 2\Gamma_\text{out}$ indicating the spin degeneracy of the system. Characterisation of tunnelling rates, electron temperature and lever arm is presented in \aref{B}. 
	
In this Letter, we operate this device as a Szilard engine, the details of which are described in the next section. This requires us to let the quantum dot system thermalise with the reservoir while keeping its energy at $E_0 = k_\text{B}T\ln 2$ which corresponds to a $50\%-50\%$ occupation. The charge sensor measures the state, and the voltage $V_\text{G2}$ is modified to realise the optimal protocol as shown in \sfref{fig:fig1}{c,d}.
	
\emph{Szilard Engine and Optimisation.}
In this section we will present how the setup is used as a Szilard engine and then proceed to optimise it. The quantum dot can be effectively described as a two-level system where the state $\ket{0}$ ($\ket{1}$) corresponds to the dot being empty (occupied). The Hamiltonian of the system would then be $\hat H(t) = \eps(t)\hat a^\dagger \hat a$ where $\eps(t)$ is the energy gap, which is controlled by the voltage $V_\text{G2}$, and $\hat a = \ket{0}\!\bra{1}$. We take the environment to be a thermal bath at inverse temperature $\beta$. By denoting with $p(t)$ the probability of being in the excited state, the rate equation of the system becomes
\begin{equation}\label{eq:dot_dynamics}
    \dot p = \gamma[2f(\eps) - (1+f(\eps))p]~.
\end{equation}
where $\gamma = \Gamma_{\text{out}}$.
The main goal when designing a thermal engine is to find the appropriate function $\eps(t)$ so that (on average) one can extract thermodynamic work in some finite time $\tau$: $\mathcal W[\eps(t)] = -\int_0^\tau\!\!dt~ p(t)\dot\eps(t)$, where we chose the sign convention so that $\mathcal W$ is positive when energy is gained from the system. 

As opposed to typical heat engines -- which use a hot and a cold bath, a Szilard engine makes use of a single thermal bath and measurements on the working substance. Thermodynamically, the measurement acts like a zero-temperature thermal bath~\cite{Guryanova2020,Taranto2023}.
The Szilard engine cycle steps are (cf. \sfref{fig:fig1}{c,d}): 0) The energy gap starts at $E_0$ so that $p = 1/2$ and the system is always in contact with the bath. 1) Measure the occupation of the dot with the charge sensor. 2) If the outcome of the measurement at step 1 is that the dot is in the state $\ket{0}$ ($n = 0$), quickly increase $\eps$ to a large value; if instead the outcome is $\ket{1}$ ($n = 1$), quickly decrease $\eps$ to a large negative value. The magnitude of $\eps$ is limited by the charging energy such that occupation of the quantum dot is limited to $n = \{0,1\}$. 3) The energy gap is brought back towards $E_0$ in some finite time $\tau$. When measuring $\ket{0}$, the protocol uses this information so that step 2 incurs no energy cost and we are always in a position of gaining energy during step 3 since the probability of a jump occurring is non-zero. However, by Landauer's principle, at the measurement in step 1 there is an implicit cost of $k_B T\ln 2$ that will be paid when the memory storing the result is erased. 
 
We turn towards the protocol optimisation: given total cycle time $\tau$, what is the function $\eps(t)$ that maximises the functional $\mathcal W[\eps(t)]$? In general, this is a highly non-trivial problem which requires approximations to be solved, e.g. slow or fast driving limits for analytical results or by making use of numerical techniques~\cite{Rolandi2023Optimal,rolandi2023finite,Erdman2023Pareto,Rolandi2023collective}. However, the dynamics of the occupation of QD1 (\eref{eq:dot_dynamics}) are simple enough for us to obtain general expressions for the optimal protocol $\eps(t)$ without any further approximation. 

Since we have to perform cycles, the symmetry of the problem imposes the boundary conditions $\eps(0) = \eps(\tau) = E_0$. We start by calculating the work gain
\begin{equation}
    \mathcal W[\eps(t)]= (p(\tau)-p(0))E_0+\int_0^\tau\!\!dt~ \dot p(t) \eps(t)~,
\end{equation}
where $p(0) = 0$ ($1$) if we measured $\ket{0}$ ($\ket{1}$) at the start of the protocol.
By using \eref{eq:dot_dynamics}, we write $\eps(t)$ as a function of $p(t)$ and $\dot p(t)$: $\beta \eps(t) = \ln\left[\frac{2-p(t)}{p(t) + \dot p(t)} - 1\right]$, where we are using time units such that $\gamma = 1$. This allows us to write the work gain as a functional of $p(t)$ and $\dot p(t)$:
\begin{equation}\label{eq:work_lagrangian}
    \mathcal W[p(t),\dot p(t)] = p(\tau)E_0 + k_B T\!\!\int_0^\tau\!\!dt~ \mathcal L[p(t),\dot p(t), t]~,
\end{equation}
where $\mathcal L[p,\dot p, t] = \dot p\ln\!\left[\frac{2-p}{p + \dot p} - 1\right]$. Since $\partial\mathcal L/\partial t = 0$, the Euler-Lagrange equation to obtain the function $p(t)$ that extremises work can be written as $\dot p^2\frac{2-p}{(p+\dot p)(2-2p-\dot p)} = K$, where $K := \mathcal L -\dot p\frac{\partial \mathcal L}{\partial\dot p}$ is a constant for the solution of the Euler-Lagrange equation. We solve for $\dot p$ to obtain
\begin{equation}\label{eq:opt_pdot}
    \dot p = \frac{1}{2}\frac{K(2-3p) + \sqrt{\Delta}}{2-p+K}~,
\end{equation}
where $\Delta = K^2(2-p)^2 + 8Kp(1-p)(2-p)$. We assumed $\dot p > 0$ by restricting ourselves to the scenario where we measured the state to be $\ket{0}$ at $t=0$; the $\ket{1}$ case is completely analogous.

We solve this differential equation to find the solution
\begin{equation}\label{eq:time_p}
    t = \int_{0}^{p(t)}\!\!dp\frac{4-2p}{K(2-3p) + \sqrt{\Delta}} ~.
\end{equation}
We therefore have an implicit formula for the optimal probability trajectory as a function of time and the integration constant: $p(t) = F_K^{-1}(t)$, where we define $F_K$ to be the solution of the integral in \eref{eq:time_p}.

\begin{figure*}[ht]
    \centering
    \includegraphics[width=\textwidth]{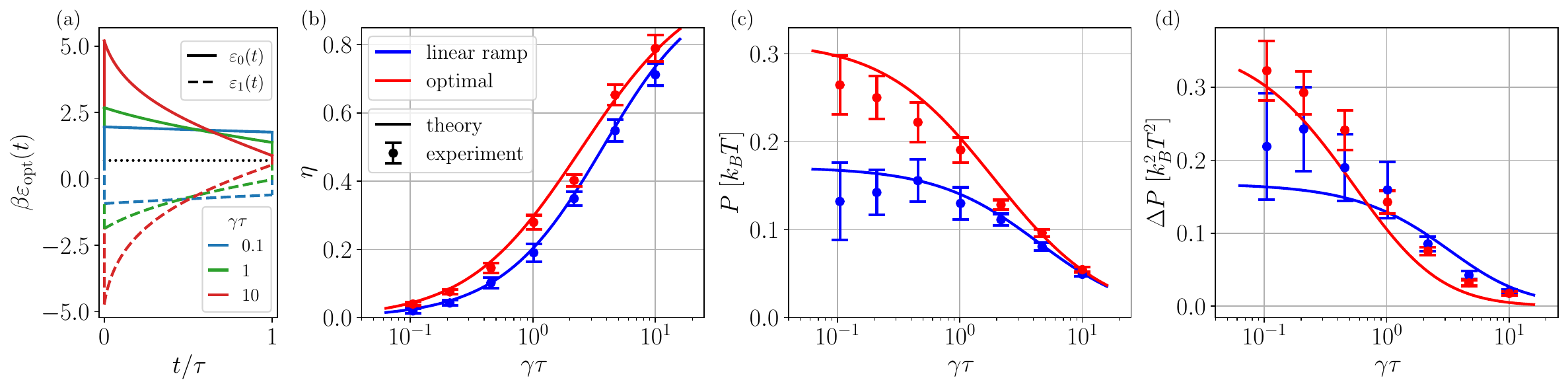}
    \vspace*{-20pt}
    \caption{(a) Optimal protocol for different values of $\gamma\tau$. The solid (dashed) lines represent the case where the state $\ket{0}$ ($\ket{1}$) is measured at the start of the cycle. The black dotted line corresponds to the initial and final value $\beta E_0 = \log(2)$ for both protocols. (b, c, d) Comparison of predicted and measured efficiency $\eta = \beta \mathcal W/\ln 2$ (b), power $P = \mathcal W/\tau$ (c), and power fluctuations $\Delta P = \text{Var}(P)$ (d) for implementations of the optimal protocol \eref{eq:opt_eps} (red) and a linear ramp $\eps(t) = E_0 + 5k_B T(\tau-t)/\tau$ (blue) at different values of $\gamma\tau$ that range from the fast-driving regime to the slow-driving regime.}
    \vspace*{-20pt}
    \label{fig:main}
\end{figure*}

Combining \eref{eq:opt_pdot} and \eref{eq:work_lagrangian}, we obtain a formula for the maximal amount of extracted work as a function of the boundary conditions and the integration constant $K$:
\begin{equation}\label{eq:min_work}
    \max_{\eps(t)}\mathcal W[\eps(t)] = p(\tau)E_0 + k_BTG_K(p(\tau))~,
\end{equation}
with
\begin{equation}
    G_K(P) :=\int_{0}^{P}\!\!\!dp\ln\!\left[\frac{2(2-p)(2-p+K)}{(2-p)(K+2p)+\sqrt{\Delta}}-1\right]~.
\end{equation}
The measurement at step 1 sets the boundary condition $p(0) = 0$. Since at the start of the next cycle another measurement will be performed, we do not need to impose a boundary condition at $p(\tau)$. By replacing $p(\tau)$ with $F_K^{-1}(\tau)$ in \eref{eq:min_work} and maximising with respect to $K$, we find the optimal integration constant $\kappa_\tau$ (for a given protocol time $\tau$) which defines the optimal protocol:
\begin{equation}\label{eq:opt_eps}
    \beta\eps_0(t) = \left.\ln\!\left[\frac{2(2-p)(2-p+K)}{(2-p)(K+2p)+\sqrt{\Delta}}-1\right]\right|_{p=F_{\kappa_\tau}^{-1}(t)}.
\end{equation}
Since the optimal integration constant is defined by
\begin{equation}
    \kappa_\tau := \argmin_{K} F_K^{-1}(\tau)E_0 + G(F_K^{-1}(\tau))~,
\end{equation}
we turned the functional minimisation problem in \eref{eq:work_lagrangian} into a regular minimisation problem, which is much simpler to handle numerically. Note from \eref{eq:opt_eps} that $\eps(0) > E_0$, showing that the optimal protocols features jumps at the start and the end of the protocol. In \sfref{fig:main}{a} we showcase these optimal protocols for a range of values of $\gamma\tau$ from the slow-driving regime to the fast-driving regime. One can note the qualitative difference of these optimal protocols at the opposite ends of the spectrum of driving speeds. In the fast driving regime, optimal protocols feature very little continuous control and consist of mainly one jump at the start and one at the end -- these protocols are also known as ``bang-bang protocols''~\cite{Rolandi2023Optimal}. However, in the slow driving regime, the optimal protocols do not feature a jump at the end and consist of a smooth and continuous driving of the system's energy, which matches with previous results in the literature~\cite{Abiuso2020}.

\emph{Results and discussion.}
For the engine cycle to be truly closed, the information obtained from the measurement in step 1 will eventually have to be erased, thus dissipating at least $k_BT\ln 2$ of heat into the environment because of Landauer's principle. This gives us a simple formula for the efficiency of this engine
\begin{equation}
    \eta = \frac{\mathcal W}{k_BT\ln 2}~.
\end{equation}
The efficiency of the information engine reaches its maximum $\eta_C = 1$ in the static limit, which corresponds to the Carnot efficiency that one obtains when setting the temperature of the cold bath to zero. This is one of many examples linking perfect measurements to zero-temperature baths~\cite{Guryanova2020,Taranto2023}. 

It is interesting to note that this expression for efficiency implies that, for a given cycle length $\tau$, the optimisation of power and efficiency coincide. It is therefore sufficient to maximise the work gained to optimise both the power and efficiency. In \sfref{fig:main}{b, c} we show as red solid lines the maximal efficiency and maximum power that can be achieved for a given value of $\gamma\tau$, and as dots with error-bars the measured work extracted from in the experimental realisation. The error is computed as the uncertainty statistical uncertainty of the measured work (cf. \aref{D}). Since the occupation is measured continuously throughout the implementation of the protocol, we used the occupation to compute the work cost of a single round of the experiment, with multiple repetitions -- in the thousands for the faster protocols and at least $200$ for the slower ones (fewer are required as the fluctuations are smaller). 

Further optimisation of power or efficiency can be achieved by tuning the cycle duration $\tau$. Indeed, we can see that in the slow driving regime the efficiency tends to the Carnot efficiency and the power  vanishes. Conversely, in the fast driving regime the power is reaching a maximum and the efficiency is vanishing.

We compare results to a ``naive protocol'': a linear ramp from $E_0 + 5k_B T$ to $E_0$ over the whole cycle length $\tau$. We see that, in terms of its absolute value, the difference is not very significant for the efficiency. However, in the fast driving regime, despite the efficiency being small, there is a large relative gain: the ratio of the two efficiencies is $\approx 1.8$. Similarly, for power, the optimal protocol has significant gains in the fast driving regime. In \sfref{fig:main}{d} we computed and measured the power fluctuations of these protocols
\begin{equation}\label{eq:var_work}
\Delta P:=\frac{2}{\tau}\,\int^\tau_0 \!\! dt\! \int^t_0\!\! dt'~ \dot\eps(t)\dot\eps(t')p(t')(q(t|t')-p(t))~,
\end{equation}
where $q(t|t')$ is the probability of the dot being in the state $\ket{1}$ at time $t$ given that it was in state $\ket{1}$ at time $t'<t$. 

The wide range of driving speeds that we had access to with the experiment allows us to appreciate how the trade-off between power optimisation and fluctuations optimisation changes depending on the driving speed. The protocols that we implemented optimise the work extracted for a given protocol duration. Therefore, for that fixed protocol duration, the power and efficiency of the Szilard engine will be optimal. Whereas the fluctuations are not optimised for, and therefore should not be expected to be optimal. Indeed, in the fast driving regime we note that the engine has large fluctuations -- in addition to the low efficiency.
However, in the slow driving regime we recover the fluctuation-dissipation relation $\frac{\tau\beta}{2}\Delta P = -\mathcal W -\Delta F$~\cite{Weber1956,Kubo1966,Miller2019}, where $\Delta F = -k_B T\ln 2$ is the difference in free energy between the start and end of the protocol. Thus, in the slow-driving regime, maximising work equals minimising fluctuations. We therefore expect our engine to be optimal for fluctuations in this regime: in \sfref{fig:main}{d} we observe that the measured fluctuations become vanishing as we increase the protocol length.

In general, the level of agreement between experimental results and theoretical predictions is high. However, it is worth noting that there seems to be a mismatch between the measured values of fluctuations and the predicted values. This can be caused by a number of reasons, however, the most significant here is a drift in the calibration of the experiment over time, which introduces a bias in the implemented protocol. This drift is possibly due to presence of trapped charges in the device. Indeed, to accurately measure the fluctuations of work one needs a much larger number of samples compared to the number of runs needed to accurately measure the average of the work extracted. This need for a large number of samples gives an additional disadvantage: the more time is needed to perform the measurements the more apparatus' calibration drift will affect the measurements. This can be seen in \sfref{fig:main}{d} for the protocols with $1 < \gamma\tau < 10$, as the repeated implementation of these left more time for the calibration to drift. Indeed, for these protocols we see that the predicted value for fluctuations is in a larger disagreement with the measurement compared to the fast protocols with $0.1 < \gamma\tau < 1$.
However, such a large disagreement is not observed for the predicted power and efficiency, even if they are computed from the same data that experienced the same calibration drift. The mismatch occurs because the protocols remain near-optimal despite drift. Therefore, for small deviation $\delta\eps$ from the optimal protocol, the power and efficiency will be affected by a factor proportional to $\delta\eps^2$, since the optimal protocol maximizes power and efficiency. Conversely, the fluctuations are not optimised by the protocols being implemented, meaning the deviations from the protocol will affect them by a factor proportional to $\delta\eps$ instead of $\delta\eps^2$. This matches \fref{fig:main}, where deviations are larger for fluctuations than for power and efficiency. 
In \aref{C} we explore this further by analysing quantitatively how much the power and fluctuations change for a constant shift from the optimal protocol $\eps(t) \rightarrow \eps(t) + \Delta$, finding very good agreement with the qualitative description provided above.

It is worth noting that this difference in scaling can be exploited to trade a small loss in power and efficiency for a large reduction in fluctuations.

\emph{Conclusion.}
We implemented and optimised a Szilard engine, where information gained through measurement allows for the extraction of work in the presence of a single heat bath. This represents a minimal model to explore the extraction of work from thermal fluctuations. 
The experimental setup consisted of a quantum dot system in a germanium quantum well, where the occupancy of the dot is manipulated and monitored. Using this system, we successfully implemented an optimised finite-time Szilard engine over two decades of driving speed ($10^{-1}\leq\gamma\tau \leq 10^1$), spanning both fast- and slow-driving regimes.
   
Our optimisation procedure maximises the extracted energy for a given, arbitrary, cycle length. In our system, this allows for the simultaneous optimisation of power and efficiency. These optimal protocols outperform a naive ramp in both power and efficiency, especially in the fast regime. We also observed that higher power inevitably leads to greater fluctuations. However, we also observed that, by slightly deviating from the power-optimal protocol, one can accept a small reduction in power in exchange for a large reduction in fluctuations thanks to the fact that we are operating close to optimum for power.

The experimental results corroborated the theoretical predictions, showing a high degree of precision in the (indirect) measurement of extracted work. However, for the fluctuations, it seems that the measured values are in larger disagreement with the theory than the corresponding measurements of work (more than $1\sigma$ away for most values). This bias can be explained by the fact that the calibration of the experiment drifts over time. Despite this, the overall agreement between theory and experiment for the extracted work demonstrates the feasibility of implementing optimally controlled information engines in solid-state platforms. This work enables studies of finite-time thermodynamics in quantum devices, such as of collective phenomena~\cite{Rolandi2023collective}, or to evaluate work fluctuations and their optimization~\cite{Denzler2024,Denzler2021,Rolandi2023Optimal}.\\

\emph{The data that support the findings of this article are openly available \cite{kushdata}}.

\begin{acknowledgments}
\vspace{-7pt}
We thank Georgios Katsaros for providing the device for this experiment. K.A. and N.A. acknowledge the support provided by funding from the Engineering and Physical Sciences Research Council IAA (Grant number EP/X525777/1).
N.A. acknowledges support from the European Research Council (grant agreement 948932) and the Royal Society (URF-R1-191150). 
A.R. is supported by the Swiss National Science Foundation through a Postdoc.Mobility (Grant No. P500PT 225461).  M.T.M. is supported by a Royal Society University Research Fellowship (URF\textbackslash R1\textbackslash 221571). M.P.-L. acknowledges support from the Spanish Agencia Estatal de Investigacion through the grant  ``Ram{\'o}n y Cajal RYC2022-036958-I''. This project is co-funded by the European Union and UK Research \& Innovation (Quantum Flagship project ASPECTS, Grant Agreement No.~101080167). Views and opinions expressed are however those of the authors only and do not necessarily reflect those of the European Union, Research Executive Agency or UK Research \& Innovation. Neither the European Union nor UK Research \& Innovation can be held responsible for them.
\end{acknowledgments}

\bibliography{references.bib}

\begin{thebibliography}{47}%
\makeatletter
\providecommand \@ifxundefined [1]{%
 \@ifx{#1\undefined}
}%
\providecommand \@ifnum [1]{%
 \ifnum #1\expandafter \@firstoftwo
 \else \expandafter \@secondoftwo
 \fi
}%
\providecommand \@ifx [1]{%
 \ifx #1\expandafter \@firstoftwo
 \else \expandafter \@secondoftwo
 \fi
}%
\providecommand \natexlab [1]{#1}%
\providecommand \enquote  [1]{``#1''}%
\providecommand \bibnamefont  [1]{#1}%
\providecommand \bibfnamefont [1]{#1}%
\providecommand \citenamefont [1]{#1}%
\providecommand \href@noop [0]{\@secondoftwo}%
\providecommand \href [0]{\begingroup \@sanitize@url \@href}%
\providecommand \@href[1]{\@@startlink{#1}\@@href}%
\providecommand \@@href[1]{\endgroup#1\@@endlink}%
\providecommand \@sanitize@url [0]{\catcode `\\12\catcode `\$12\catcode `\&12\catcode `\#12\catcode `\^12\catcode `\_12\catcode `\%12\relax}%
\providecommand \@@startlink[1]{}%
\providecommand \@@endlink[0]{}%
\providecommand \url  [0]{\begingroup\@sanitize@url \@url }%
\providecommand \@url [1]{\endgroup\@href {#1}{\urlprefix }}%
\providecommand \urlprefix  [0]{URL }%
\providecommand \Eprint [0]{\href }%
\providecommand \doibase [0]{https://doi.org/}%
\providecommand \selectlanguage [0]{\@gobble}%
\providecommand \bibinfo  [0]{\@secondoftwo}%
\providecommand \bibfield  [0]{\@secondoftwo}%
\providecommand \translation [1]{[#1]}%
\providecommand \BibitemOpen [0]{}%
\providecommand \bibitemStop [0]{}%
\providecommand \bibitemNoStop [0]{.\EOS\space}%
\providecommand \EOS [0]{\spacefactor3000\relax}%
\providecommand \BibitemShut  [1]{\csname bibitem#1\endcsname}%
\let\auto@bib@innerbib\@empty
\bibitem [{\citenamefont {Shiraishi}\ \emph {et~al.}(2016)\citenamefont {Shiraishi}, \citenamefont {Saito},\ and\ \citenamefont {Tasaki}}]{Shiraishi2016}%
  \BibitemOpen
  \bibfield  {author} {\bibinfo {author} {\bibfnamefont {N.}~\bibnamefont {Shiraishi}}, \bibinfo {author} {\bibfnamefont {K.}~\bibnamefont {Saito}},\ and\ \bibinfo {author} {\bibfnamefont {H.}~\bibnamefont {Tasaki}},\ }\bibfield  {title} {\bibinfo {title} {Universal trade-off relation between power and efficiency for heat engines},\ }\href {https://doi.org/10.1103/PhysRevLett.117.190601} {\bibfield  {journal} {\bibinfo  {journal} {Phys. Rev. Lett.}\ }\textbf {\bibinfo {volume} {117}},\ \bibinfo {pages} {190601} (\bibinfo {year} {2016})}\BibitemShut {NoStop}%
\bibitem [{\citenamefont {Campisi}\ and\ \citenamefont {Fazio}(2016)}]{Campisi2016}%
  \BibitemOpen
  \bibfield  {author} {\bibinfo {author} {\bibfnamefont {M.}~\bibnamefont {Campisi}}\ and\ \bibinfo {author} {\bibfnamefont {R.}~\bibnamefont {Fazio}},\ }\bibfield  {title} {\bibinfo {title} {The power of a critical heat engine},\ }\href {https://doi.org/10.1038/ncomms11895} {\bibfield  {journal} {\bibinfo  {journal} {Nature Communications}\ }\textbf {\bibinfo {volume} {7}},\ \bibinfo {pages} {11895} (\bibinfo {year} {2016})}\BibitemShut {NoStop}%
\bibitem [{\citenamefont {Pietzonka}\ and\ \citenamefont {Seifert}(2018)}]{Pietzonka2018}%
  \BibitemOpen
  \bibfield  {author} {\bibinfo {author} {\bibfnamefont {P.}~\bibnamefont {Pietzonka}}\ and\ \bibinfo {author} {\bibfnamefont {U.}~\bibnamefont {Seifert}},\ }\bibfield  {title} {\bibinfo {title} {Universal trade-off between power, efficiency, and constancy in steady-state heat engines},\ }\href {https://doi.org/10.1103/PhysRevLett.120.190602} {\bibfield  {journal} {\bibinfo  {journal} {Phys. Rev. Lett.}\ }\textbf {\bibinfo {volume} {120}},\ \bibinfo {pages} {190602} (\bibinfo {year} {2018})}\BibitemShut {NoStop}%
\bibitem [{\citenamefont {Szilard}(1964)}]{szilard1964decrease}%
  \BibitemOpen
  \bibfield  {author} {\bibinfo {author} {\bibfnamefont {L.}~\bibnamefont {Szilard}},\ }\bibfield  {title} {\bibinfo {title} {On the decrease of entropy in a thermodynamic system by the intervention of intelligent beings},\ }\href@noop {} {\bibfield  {journal} {\bibinfo  {journal} {Behavioral Science}\ }\textbf {\bibinfo {volume} {9}},\ \bibinfo {pages} {301} (\bibinfo {year} {1964})}\BibitemShut {NoStop}%
\bibitem [{\citenamefont {Landauer}(1991)}]{Landauer1991}%
  \BibitemOpen
  \bibfield  {author} {\bibinfo {author} {\bibfnamefont {R.}~\bibnamefont {Landauer}},\ }\bibfield  {title} {\bibinfo {title} {Information is physical},\ }\href {https://doi.org/10.1063/1.881299} {\bibfield  {journal} {\bibinfo  {journal} {Physics Today}\ }\textbf {\bibinfo {volume} {44}},\ \bibinfo {pages} {23–29} (\bibinfo {year} {1991})}\BibitemShut {NoStop}%
\bibitem [{\citenamefont {Parrondo}\ \emph {et~al.}(2015)\citenamefont {Parrondo}, \citenamefont {Horowitz},\ and\ \citenamefont {Sagawa}}]{Parrondo2015}%
  \BibitemOpen
  \bibfield  {author} {\bibinfo {author} {\bibfnamefont {J.~M.~R.}\ \bibnamefont {Parrondo}}, \bibinfo {author} {\bibfnamefont {J.~M.}\ \bibnamefont {Horowitz}},\ and\ \bibinfo {author} {\bibfnamefont {T.}~\bibnamefont {Sagawa}},\ }\bibfield  {title} {\bibinfo {title} {Thermodynamics of information},\ }\href {https://doi.org/10.1038/nphys3230} {\bibfield  {journal} {\bibinfo  {journal} {Nature Physics}\ }\textbf {\bibinfo {volume} {11}},\ \bibinfo {pages} {131–139} (\bibinfo {year} {2015})}\BibitemShut {NoStop}%
\bibitem [{\citenamefont {du~Buisson}\ \emph {et~al.}(2024)\citenamefont {du~Buisson}, \citenamefont {Sivak},\ and\ \citenamefont {Bechhoefer}}]{duBuisson2024}%
  \BibitemOpen
  \bibfield  {author} {\bibinfo {author} {\bibfnamefont {J.}~\bibnamefont {du~Buisson}}, \bibinfo {author} {\bibfnamefont {D.~A.}\ \bibnamefont {Sivak}},\ and\ \bibinfo {author} {\bibfnamefont {J.}~\bibnamefont {Bechhoefer}},\ }\bibfield  {title} {\bibinfo {title} {Performance limits of information engines},\ }\href {https://doi.org/10.1080/23746149.2024.2352112} {\bibfield  {journal} {\bibinfo  {journal} {Advances in Physics: X}\ }\textbf {\bibinfo {volume} {9}} (\bibinfo {year} {2024})}\BibitemShut {NoStop}%
\bibitem [{\citenamefont {Josefsson}\ \emph {et~al.}(2018)\citenamefont {Josefsson}, \citenamefont {Svilans}, \citenamefont {Burke}, \citenamefont {Hoffmann}, \citenamefont {Fahlvik}, \citenamefont {Thelander}, \citenamefont {Leijnse},\ and\ \citenamefont {Linke}}]{Josefsson2018}%
  \BibitemOpen
  \bibfield  {author} {\bibinfo {author} {\bibfnamefont {M.}~\bibnamefont {Josefsson}}, \bibinfo {author} {\bibfnamefont {A.}~\bibnamefont {Svilans}}, \bibinfo {author} {\bibfnamefont {A.~M.}\ \bibnamefont {Burke}}, \bibinfo {author} {\bibfnamefont {E.~A.}\ \bibnamefont {Hoffmann}}, \bibinfo {author} {\bibfnamefont {S.}~\bibnamefont {Fahlvik}}, \bibinfo {author} {\bibfnamefont {C.}~\bibnamefont {Thelander}}, \bibinfo {author} {\bibfnamefont {M.}~\bibnamefont {Leijnse}},\ and\ \bibinfo {author} {\bibfnamefont {H.}~\bibnamefont {Linke}},\ }\bibfield  {title} {\bibinfo {title} {A quantum-dot heat engine operating close to the thermodynamic efficiency limits},\ }\href {https://doi.org/10.1038/s41565-018-0200-5} {\bibfield  {journal} {\bibinfo  {journal} {Nature Nanotechnology}\ }\textbf {\bibinfo {volume} {13}},\ \bibinfo {pages} {920} (\bibinfo {year} {2018})}\BibitemShut {NoStop}%
\bibitem [{\citenamefont {Karimi}\ \emph {et~al.}(2020)\citenamefont {Karimi}, \citenamefont {Brange}, \citenamefont {Samuelsson},\ and\ \citenamefont {Pekola}}]{Karimi2020}%
  \BibitemOpen
  \bibfield  {author} {\bibinfo {author} {\bibfnamefont {B.}~\bibnamefont {Karimi}}, \bibinfo {author} {\bibfnamefont {F.}~\bibnamefont {Brange}}, \bibinfo {author} {\bibfnamefont {P.}~\bibnamefont {Samuelsson}},\ and\ \bibinfo {author} {\bibfnamefont {J.~P.}\ \bibnamefont {Pekola}},\ }\bibfield  {title} {\bibinfo {title} {Reaching the ultimate energy resolution of a quantum detector},\ }\href {https://doi.org/10.1038/s41467-019-14247-2} {\bibfield  {journal} {\bibinfo  {journal} {Nature Communications}\ }\textbf {\bibinfo {volume} {11}},\ \bibinfo {pages} {367} (\bibinfo {year} {2020})}\BibitemShut {NoStop}%
\bibitem [{\citenamefont {Majidi}\ \emph {et~al.}(2022)\citenamefont {Majidi}, \citenamefont {Josefsson}, \citenamefont {Kumar}, \citenamefont {Leijnse}, \citenamefont {Samuelson}, \citenamefont {Courtois}, \citenamefont {Winkelmann},\ and\ \citenamefont {Maisi}}]{Majidi2022}%
  \BibitemOpen
  \bibfield  {author} {\bibinfo {author} {\bibfnamefont {D.}~\bibnamefont {Majidi}}, \bibinfo {author} {\bibfnamefont {M.}~\bibnamefont {Josefsson}}, \bibinfo {author} {\bibfnamefont {M.}~\bibnamefont {Kumar}}, \bibinfo {author} {\bibfnamefont {M.}~\bibnamefont {Leijnse}}, \bibinfo {author} {\bibfnamefont {L.}~\bibnamefont {Samuelson}}, \bibinfo {author} {\bibfnamefont {H.}~\bibnamefont {Courtois}}, \bibinfo {author} {\bibfnamefont {C.~B.}\ \bibnamefont {Winkelmann}},\ and\ \bibinfo {author} {\bibfnamefont {V.~F.}\ \bibnamefont {Maisi}},\ }\bibfield  {title} {\bibinfo {title} {Quantum {{Confinement Suppressing Electronic Heat Flow}} below the {{Wiedemann}}--{{Franz Law}}},\ }\href {https://doi.org/10.1021/acs.nanolett.1c03437} {\bibfield  {journal} {\bibinfo  {journal} {Nano Letters}\ }\textbf {\bibinfo {volume} {22}},\ \bibinfo {pages} {630} (\bibinfo {year} {2022})}\BibitemShut {NoStop}%
\bibitem [{\citenamefont {Majidi}\ \emph {et~al.}(2024)\citenamefont {Majidi}, \citenamefont {Bergfield}, \citenamefont {Maisi}, \citenamefont {H{\"o}fer}, \citenamefont {Courtois},\ and\ \citenamefont {Winkelmann}}]{Majidi2024}%
  \BibitemOpen
  \bibfield  {author} {\bibinfo {author} {\bibfnamefont {D.}~\bibnamefont {Majidi}}, \bibinfo {author} {\bibfnamefont {J.~P.}\ \bibnamefont {Bergfield}}, \bibinfo {author} {\bibfnamefont {V.}~\bibnamefont {Maisi}}, \bibinfo {author} {\bibfnamefont {J.}~\bibnamefont {H{\"o}fer}}, \bibinfo {author} {\bibfnamefont {H.}~\bibnamefont {Courtois}},\ and\ \bibinfo {author} {\bibfnamefont {C.~B.}\ \bibnamefont {Winkelmann}},\ }\bibfield  {title} {\bibinfo {title} {Heat transport at the nanoscale and ultralow temperatures---{{Implications}} for quantum technologies},\ }\href {https://doi.org/10.1063/5.0204207} {\bibfield  {journal} {\bibinfo  {journal} {Applied Physics Letters}\ }\textbf {\bibinfo {volume} {124}},\ \bibinfo {pages} {140504} (\bibinfo {year} {2024})}\BibitemShut {NoStop}%
\bibitem [{\citenamefont {Maxwell}(2011)}]{Maxwell11}%
  \BibitemOpen
  \bibfield  {author} {\bibinfo {author} {\bibfnamefont {J.~C.}\ \bibnamefont {Maxwell}},\ }\href {https://doi.org/10.1017/cbo9781139057943} {\emph {\bibinfo {title} {Theory of Heat}}}\ (\bibinfo  {publisher} {Cambridge University Press},\ \bibinfo {year} {2011})\BibitemShut {NoStop}%
\bibitem [{\citenamefont {Landauer}(1961)}]{landauer1961irreversibility}%
  \BibitemOpen
  \bibfield  {author} {\bibinfo {author} {\bibfnamefont {R.}~\bibnamefont {Landauer}},\ }\bibfield  {title} {\bibinfo {title} {Irreversibility and heat generation in the computing process},\ }\href@noop {} {\bibfield  {journal} {\bibinfo  {journal} {IBM journal of research and development}\ }\textbf {\bibinfo {volume} {5}},\ \bibinfo {pages} {183} (\bibinfo {year} {1961})}\BibitemShut {NoStop}%
\bibitem [{\citenamefont {Ciliberto}(2017)}]{Ciliberto2017}%
  \BibitemOpen
  \bibfield  {author} {\bibinfo {author} {\bibfnamefont {S.}~\bibnamefont {Ciliberto}},\ }\bibfield  {title} {\bibinfo {title} {Experiments in stochastic thermodynamics: Short history and perspectives},\ }\href {https://doi.org/10.1103/PhysRevX.7.021051} {\bibfield  {journal} {\bibinfo  {journal} {Phys. Rev. X}\ }\textbf {\bibinfo {volume} {7}},\ \bibinfo {pages} {021051} (\bibinfo {year} {2017})}\BibitemShut {NoStop}%
\bibitem [{\citenamefont {Myers}\ \emph {et~al.}(2022)\citenamefont {Myers}, \citenamefont {Abah},\ and\ \citenamefont {Deffner}}]{Myers2022}%
  \BibitemOpen
  \bibfield  {author} {\bibinfo {author} {\bibfnamefont {N.~M.}\ \bibnamefont {Myers}}, \bibinfo {author} {\bibfnamefont {O.}~\bibnamefont {Abah}},\ and\ \bibinfo {author} {\bibfnamefont {S.}~\bibnamefont {Deffner}},\ }\bibfield  {title} {\bibinfo {title} {Quantum thermodynamic devices: From theoretical proposals to experimental reality},\ }\href {https://doi.org/10.1116/5.0083192} {\bibfield  {journal} {\bibinfo  {journal} {AVS Quantum Science}\ }\textbf {\bibinfo {volume} {4}} (\bibinfo {year} {2022})}\BibitemShut {NoStop}%
\bibitem [{\citenamefont {Toyabe}\ \emph {et~al.}(2010)\citenamefont {Toyabe}, \citenamefont {Sagawa}, \citenamefont {Ueda}, \citenamefont {Muneyuki},\ and\ \citenamefont {Sano}}]{Toyabe2010}%
  \BibitemOpen
  \bibfield  {author} {\bibinfo {author} {\bibfnamefont {S.}~\bibnamefont {Toyabe}}, \bibinfo {author} {\bibfnamefont {T.}~\bibnamefont {Sagawa}}, \bibinfo {author} {\bibfnamefont {M.}~\bibnamefont {Ueda}}, \bibinfo {author} {\bibfnamefont {E.}~\bibnamefont {Muneyuki}},\ and\ \bibinfo {author} {\bibfnamefont {M.}~\bibnamefont {Sano}},\ }\bibfield  {title} {\bibinfo {title} {Experimental demonstration of information-to-energy conversion and validation of the generalized jarzynski equality},\ }\href {https://doi.org/10.1038/nphys1821} {\bibfield  {journal} {\bibinfo  {journal} {Nature Physics}\ }\textbf {\bibinfo {volume} {6}},\ \bibinfo {pages} {988–992} (\bibinfo {year} {2010})}\BibitemShut {NoStop}%
\bibitem [{\citenamefont {Bérut}\ \emph {et~al.}(2012)\citenamefont {Bérut}, \citenamefont {Arakelyan}, \citenamefont {Petrosyan}, \citenamefont {Ciliberto}, \citenamefont {Dillenschneider},\ and\ \citenamefont {Lutz}}]{Brut2012}%
  \BibitemOpen
  \bibfield  {author} {\bibinfo {author} {\bibfnamefont {A.}~\bibnamefont {Bérut}}, \bibinfo {author} {\bibfnamefont {A.}~\bibnamefont {Arakelyan}}, \bibinfo {author} {\bibfnamefont {A.}~\bibnamefont {Petrosyan}}, \bibinfo {author} {\bibfnamefont {S.}~\bibnamefont {Ciliberto}}, \bibinfo {author} {\bibfnamefont {R.}~\bibnamefont {Dillenschneider}},\ and\ \bibinfo {author} {\bibfnamefont {E.}~\bibnamefont {Lutz}},\ }\bibfield  {title} {\bibinfo {title} {Experimental verification of landauer’s principle linking information and thermodynamics},\ }\href {https://doi.org/10.1038/nature10872} {\bibfield  {journal} {\bibinfo  {journal} {Nature}\ }\textbf {\bibinfo {volume} {483}},\ \bibinfo {pages} {187–189} (\bibinfo {year} {2012})}\BibitemShut {NoStop}%
\bibitem [{\citenamefont {Jun}\ \emph {et~al.}(2014)\citenamefont {Jun}, \citenamefont {Gavrilov},\ and\ \citenamefont {Bechhoefer}}]{Jun2014High}%
  \BibitemOpen
  \bibfield  {author} {\bibinfo {author} {\bibfnamefont {Y.}~\bibnamefont {Jun}}, \bibinfo {author} {\bibfnamefont {M.}~\bibnamefont {Gavrilov}},\ and\ \bibinfo {author} {\bibfnamefont {J.}~\bibnamefont {Bechhoefer}},\ }\bibfield  {title} {\bibinfo {title} {High-precision test of landauer's principle in a feedback trap},\ }\href {https://doi.org/10.1103/PhysRevLett.113.190601} {\bibfield  {journal} {\bibinfo  {journal} {Phys. Rev. Lett.}\ }\textbf {\bibinfo {volume} {113}},\ \bibinfo {pages} {190601} (\bibinfo {year} {2014})}\BibitemShut {NoStop}%
\bibitem [{\citenamefont {Paneru}\ \emph {et~al.}(2018)\citenamefont {Paneru}, \citenamefont {Lee}, \citenamefont {Tlusty},\ and\ \citenamefont {Pak}}]{Paneru2018}%
  \BibitemOpen
  \bibfield  {author} {\bibinfo {author} {\bibfnamefont {G.}~\bibnamefont {Paneru}}, \bibinfo {author} {\bibfnamefont {D.~Y.}\ \bibnamefont {Lee}}, \bibinfo {author} {\bibfnamefont {T.}~\bibnamefont {Tlusty}},\ and\ \bibinfo {author} {\bibfnamefont {H.~K.}\ \bibnamefont {Pak}},\ }\bibfield  {title} {\bibinfo {title} {Lossless brownian information engine},\ }\href {https://doi.org/10.1103/PhysRevLett.120.020601} {\bibfield  {journal} {\bibinfo  {journal} {Phys. Rev. Lett.}\ }\textbf {\bibinfo {volume} {120}},\ \bibinfo {pages} {020601} (\bibinfo {year} {2018})}\BibitemShut {NoStop}%
\bibitem [{\citenamefont {Admon}\ \emph {et~al.}(2018)\citenamefont {Admon}, \citenamefont {Rahav},\ and\ \citenamefont {Roichman}}]{Admon2018}%
  \BibitemOpen
  \bibfield  {author} {\bibinfo {author} {\bibfnamefont {T.}~\bibnamefont {Admon}}, \bibinfo {author} {\bibfnamefont {S.}~\bibnamefont {Rahav}},\ and\ \bibinfo {author} {\bibfnamefont {Y.}~\bibnamefont {Roichman}},\ }\bibfield  {title} {\bibinfo {title} {Experimental realization of an information machine with tunable temporal correlations},\ }\href {https://doi.org/10.1103/PhysRevLett.121.180601} {\bibfield  {journal} {\bibinfo  {journal} {Phys. Rev. Lett.}\ }\textbf {\bibinfo {volume} {121}},\ \bibinfo {pages} {180601} (\bibinfo {year} {2018})}\BibitemShut {NoStop}%
\bibitem [{\citenamefont {Koski}\ \emph {et~al.}(2014)\citenamefont {Koski}, \citenamefont {Maisi}, \citenamefont {Pekola},\ and\ \citenamefont {Averin}}]{koski_experimental_2014}%
  \BibitemOpen
  \bibfield  {author} {\bibinfo {author} {\bibfnamefont {J.~V.}\ \bibnamefont {Koski}}, \bibinfo {author} {\bibfnamefont {V.~F.}\ \bibnamefont {Maisi}}, \bibinfo {author} {\bibfnamefont {J.~P.}\ \bibnamefont {Pekola}},\ and\ \bibinfo {author} {\bibfnamefont {D.~V.}\ \bibnamefont {Averin}},\ }\bibfield  {title} {\bibinfo {title} {Experimental realization of a {Szilard} engine with a single electron},\ }\href {https://doi.org/10.1073/pnas.1406966111} {\bibfield  {journal} {\bibinfo  {journal} {PNAS}\ }\textbf {\bibinfo {volume} {111}},\ \bibinfo {pages} {13786} (\bibinfo {year} {2014})}\BibitemShut {NoStop}%
\bibitem [{\citenamefont {Barker}\ \emph {et~al.}(2022)\citenamefont {Barker}, \citenamefont {Scandi}, \citenamefont {Lehmann}, \citenamefont {Thelander}, \citenamefont {Dick}, \citenamefont {Perarnau-Llobet},\ and\ \citenamefont {Maisi}}]{Barker2022}%
  \BibitemOpen
  \bibfield  {author} {\bibinfo {author} {\bibfnamefont {D.}~\bibnamefont {Barker}}, \bibinfo {author} {\bibfnamefont {M.}~\bibnamefont {Scandi}}, \bibinfo {author} {\bibfnamefont {S.}~\bibnamefont {Lehmann}}, \bibinfo {author} {\bibfnamefont {C.}~\bibnamefont {Thelander}}, \bibinfo {author} {\bibfnamefont {K.~A.}\ \bibnamefont {Dick}}, \bibinfo {author} {\bibfnamefont {M.}~\bibnamefont {Perarnau-Llobet}},\ and\ \bibinfo {author} {\bibfnamefont {V.~F.}\ \bibnamefont {Maisi}},\ }\bibfield  {title} {\bibinfo {title} {Experimental verification of the work fluctuation-dissipation relation for information-to-work conversion},\ }\href {https://doi.org/10.1103/PhysRevLett.128.040602} {\bibfield  {journal} {\bibinfo  {journal} {Phys. Rev. Lett.}\ }\textbf {\bibinfo {volume} {128}},\ \bibinfo {pages} {040602} (\bibinfo {year} {2022})}\BibitemShut {NoStop}%
\bibitem [{\citenamefont {Scandi}\ \emph {et~al.}(2022)\citenamefont {Scandi}, \citenamefont {Barker}, \citenamefont {Lehmann}, \citenamefont {Dick}, \citenamefont {Maisi},\ and\ \citenamefont {Perarnau-Llobet}}]{Scandi2022}%
  \BibitemOpen
  \bibfield  {author} {\bibinfo {author} {\bibfnamefont {M.}~\bibnamefont {Scandi}}, \bibinfo {author} {\bibfnamefont {D.}~\bibnamefont {Barker}}, \bibinfo {author} {\bibfnamefont {S.}~\bibnamefont {Lehmann}}, \bibinfo {author} {\bibfnamefont {K.~A.}\ \bibnamefont {Dick}}, \bibinfo {author} {\bibfnamefont {V.~F.}\ \bibnamefont {Maisi}},\ and\ \bibinfo {author} {\bibfnamefont {M.}~\bibnamefont {Perarnau-Llobet}},\ }\bibfield  {title} {\bibinfo {title} {Minimally dissipative information erasure in a quantum dot via thermodynamic length},\ }\href {https://doi.org/10.1103/PhysRevLett.129.270601} {\bibfield  {journal} {\bibinfo  {journal} {Phys. Rev. Lett.}\ }\textbf {\bibinfo {volume} {129}},\ \bibinfo {pages} {270601} (\bibinfo {year} {2022})}\BibitemShut {NoStop}%
\bibitem [{\citenamefont {Kumar}\ \emph {et~al.}(2018)\citenamefont {Kumar}, \citenamefont {Wu}, \citenamefont {Giraldo},\ and\ \citenamefont {Weiss}}]{Kumar2018}%
  \BibitemOpen
  \bibfield  {author} {\bibinfo {author} {\bibfnamefont {A.}~\bibnamefont {Kumar}}, \bibinfo {author} {\bibfnamefont {T.-Y.}\ \bibnamefont {Wu}}, \bibinfo {author} {\bibfnamefont {F.}~\bibnamefont {Giraldo}},\ and\ \bibinfo {author} {\bibfnamefont {D.~S.}\ \bibnamefont {Weiss}},\ }\bibfield  {title} {\bibinfo {title} {Sorting ultracold atoms in a three-dimensional optical lattice in a realization of maxwell’s demon},\ }\href {https://doi.org/10.1038/s41586-018-0458-7} {\bibfield  {journal} {\bibinfo  {journal} {Nature}\ }\textbf {\bibinfo {volume} {561}},\ \bibinfo {pages} {83–87} (\bibinfo {year} {2018})}\BibitemShut {NoStop}%
\bibitem [{\citenamefont {Camati}\ \emph {et~al.}(2016)\citenamefont {Camati}, \citenamefont {Peterson}, \citenamefont {Batalhão}, \citenamefont {Micadei}, \citenamefont {Souza}, \citenamefont {Sarthour}, \citenamefont {Oliveira},\ and\ \citenamefont {Serra}}]{Camati2016}%
  \BibitemOpen
  \bibfield  {author} {\bibinfo {author} {\bibfnamefont {P.~A.}\ \bibnamefont {Camati}}, \bibinfo {author} {\bibfnamefont {J.~P.}\ \bibnamefont {Peterson}}, \bibinfo {author} {\bibfnamefont {T.~B.}\ \bibnamefont {Batalhão}}, \bibinfo {author} {\bibfnamefont {K.}~\bibnamefont {Micadei}}, \bibinfo {author} {\bibfnamefont {A.~M.}\ \bibnamefont {Souza}}, \bibinfo {author} {\bibfnamefont {R.~S.}\ \bibnamefont {Sarthour}}, \bibinfo {author} {\bibfnamefont {I.~S.}\ \bibnamefont {Oliveira}},\ and\ \bibinfo {author} {\bibfnamefont {R.~M.}\ \bibnamefont {Serra}},\ }\bibfield  {title} {\bibinfo {title} {Experimental rectification of entropy production by maxwell’s demon in a quantum system},\ }\href {https://doi.org/10.1103/physrevlett.117.240502} {\bibfield  {journal} {\bibinfo  {journal} {Physical Review Letters}\ }\textbf {\bibinfo {volume} {117}} (\bibinfo {year} {2016})}\BibitemShut {NoStop}%
\bibitem [{\citenamefont {Cottet}\ \emph {et~al.}(2017)\citenamefont {Cottet}, \citenamefont {Jezouin}, \citenamefont {Bretheau}, \citenamefont {Campagne-Ibarcq}, \citenamefont {Ficheux}, \citenamefont {Anders}, \citenamefont {Auffèves}, \citenamefont {Azouit}, \citenamefont {Rouchon},\ and\ \citenamefont {Huard}}]{Cottet2017}%
  \BibitemOpen
  \bibfield  {author} {\bibinfo {author} {\bibfnamefont {N.}~\bibnamefont {Cottet}}, \bibinfo {author} {\bibfnamefont {S.}~\bibnamefont {Jezouin}}, \bibinfo {author} {\bibfnamefont {L.}~\bibnamefont {Bretheau}}, \bibinfo {author} {\bibfnamefont {P.}~\bibnamefont {Campagne-Ibarcq}}, \bibinfo {author} {\bibfnamefont {Q.}~\bibnamefont {Ficheux}}, \bibinfo {author} {\bibfnamefont {J.}~\bibnamefont {Anders}}, \bibinfo {author} {\bibfnamefont {A.}~\bibnamefont {Auffèves}}, \bibinfo {author} {\bibfnamefont {R.}~\bibnamefont {Azouit}}, \bibinfo {author} {\bibfnamefont {P.}~\bibnamefont {Rouchon}},\ and\ \bibinfo {author} {\bibfnamefont {B.}~\bibnamefont {Huard}},\ }\bibfield  {title} {\bibinfo {title} {Observing a quantum maxwell demon at work},\ }\href {https://doi.org/10.1073/pnas.1704827114} {\bibfield  {journal} {\bibinfo  {journal} {Proceedings of the National Academy of Sciences}\ }\textbf {\bibinfo {volume} {114}},\ \bibinfo {pages} {7561–7564} (\bibinfo {year} {2017})}\BibitemShut {NoStop}%
\bibitem [{\citenamefont {Masuyama}\ \emph {et~al.}(2018)\citenamefont {Masuyama}, \citenamefont {Funo}, \citenamefont {Murashita}, \citenamefont {Noguchi}, \citenamefont {Kono}, \citenamefont {Tabuchi}, \citenamefont {Yamazaki}, \citenamefont {Ueda},\ and\ \citenamefont {Nakamura}}]{Masuyama2018}%
  \BibitemOpen
  \bibfield  {author} {\bibinfo {author} {\bibfnamefont {Y.}~\bibnamefont {Masuyama}}, \bibinfo {author} {\bibfnamefont {K.}~\bibnamefont {Funo}}, \bibinfo {author} {\bibfnamefont {Y.}~\bibnamefont {Murashita}}, \bibinfo {author} {\bibfnamefont {A.}~\bibnamefont {Noguchi}}, \bibinfo {author} {\bibfnamefont {S.}~\bibnamefont {Kono}}, \bibinfo {author} {\bibfnamefont {Y.}~\bibnamefont {Tabuchi}}, \bibinfo {author} {\bibfnamefont {R.}~\bibnamefont {Yamazaki}}, \bibinfo {author} {\bibfnamefont {M.}~\bibnamefont {Ueda}},\ and\ \bibinfo {author} {\bibfnamefont {Y.}~\bibnamefont {Nakamura}},\ }\bibfield  {title} {\bibinfo {title} {Information-to-work conversion by maxwell’s demon in a superconducting circuit quantum electrodynamical system},\ }\href {https://doi.org/10.1038/s41467-018-03686-y} {\bibfield  {journal} {\bibinfo  {journal} {Nature Communications}\ }\textbf {\bibinfo {volume} {9}} (\bibinfo {year} {2018})}\BibitemShut {NoStop}%
\bibitem [{\citenamefont {Esposito}\ \emph {et~al.}(2010)\citenamefont {Esposito}, \citenamefont {Kawai}, \citenamefont {Lindenberg},\ and\ \citenamefont {den Broeck}}]{Esposito2010}%
  \BibitemOpen
  \bibfield  {author} {\bibinfo {author} {\bibfnamefont {M.}~\bibnamefont {Esposito}}, \bibinfo {author} {\bibfnamefont {R.}~\bibnamefont {Kawai}}, \bibinfo {author} {\bibfnamefont {K.}~\bibnamefont {Lindenberg}},\ and\ \bibinfo {author} {\bibfnamefont {C.~V.}\ \bibnamefont {den Broeck}},\ }\bibfield  {title} {\bibinfo {title} {Finite-time thermodynamics for a single-level quantum dot},\ }\href {https://doi.org/10.1209/0295-5075/89/20003} {\bibfield  {journal} {\bibinfo  {journal} {{EPL} (Europhysics Letters)}\ }\textbf {\bibinfo {volume} {89}},\ \bibinfo {pages} {20003} (\bibinfo {year} {2010})}\BibitemShut {NoStop}%
\bibitem [{\citenamefont {Blaber}\ \emph {et~al.}(2021)\citenamefont {Blaber}, \citenamefont {Louwerse},\ and\ \citenamefont {Sivak}}]{Blaber2021Steps}%
  \BibitemOpen
  \bibfield  {author} {\bibinfo {author} {\bibfnamefont {S.}~\bibnamefont {Blaber}}, \bibinfo {author} {\bibfnamefont {M.~D.}\ \bibnamefont {Louwerse}},\ and\ \bibinfo {author} {\bibfnamefont {D.~A.}\ \bibnamefont {Sivak}},\ }\bibfield  {title} {\bibinfo {title} {Steps minimize dissipation in rapidly driven stochastic systems},\ }\href {https://doi.org/10.1103/PhysRevE.104.L022101} {\bibfield  {journal} {\bibinfo  {journal} {Phys. Rev. E}\ }\textbf {\bibinfo {volume} {104}},\ \bibinfo {pages} {L022101} (\bibinfo {year} {2021})}\BibitemShut {NoStop}%
\bibitem [{\citenamefont {Rolandi}\ \emph {et~al.}(2023{\natexlab{a}})\citenamefont {Rolandi}, \citenamefont {Perarnau-Llobet},\ and\ \citenamefont {Miller}}]{Rolandi2023Optimal}%
  \BibitemOpen
  \bibfield  {author} {\bibinfo {author} {\bibfnamefont {A.}~\bibnamefont {Rolandi}}, \bibinfo {author} {\bibfnamefont {M.}~\bibnamefont {Perarnau-Llobet}},\ and\ \bibinfo {author} {\bibfnamefont {H.~J.~D.}\ \bibnamefont {Miller}},\ }\bibfield  {title} {\bibinfo {title} {Optimal control of dissipation and work fluctuations for rapidly driven systems},\ }\href {https://doi.org/10.1088/1367-2630/ace2e3} {\bibfield  {journal} {\bibinfo  {journal} {New Journal of Physics}\ }\textbf {\bibinfo {volume} {25}},\ \bibinfo {pages} {073005} (\bibinfo {year} {2023}{\natexlab{a}})}\BibitemShut {NoStop}%
\bibitem [{\citenamefont {Scandi}\ and\ \citenamefont {{Perarnau-Llobet}}(2019)}]{scandiThermodynamicLengthOpen2019}%
  \BibitemOpen
  \bibfield  {author} {\bibinfo {author} {\bibfnamefont {M.}~\bibnamefont {Scandi}}\ and\ \bibinfo {author} {\bibfnamefont {M.}~\bibnamefont {{Perarnau-Llobet}}},\ }\bibfield  {title} {\bibinfo {title} {Thermodynamic length in open quantum systems},\ }\href {https://doi.org/10.22331/q-2019-10-24-197} {\bibfield  {journal} {\bibinfo  {journal} {Quantum}\ }\textbf {\bibinfo {volume} {3}},\ \bibinfo {pages} {197} (\bibinfo {year} {2019})}\BibitemShut {NoStop}%
\bibitem [{\citenamefont {Jarzynski}(2011)}]{Jarzynski2011}%
  \BibitemOpen
  \bibfield  {author} {\bibinfo {author} {\bibfnamefont {C.}~\bibnamefont {Jarzynski}},\ }\bibfield  {title} {\bibinfo {title} {Equalities and inequalities: Irreversibility and the second law of thermodynamics at the nanoscale},\ }\href {https://doi.org/10.1146/annurev-conmatphys-062910-140506} {\bibfield  {journal} {\bibinfo  {journal} {Annual Review of Condensed Matter Physics}\ }\textbf {\bibinfo {volume} {2}},\ \bibinfo {pages} {329} (\bibinfo {year} {2011})}\BibitemShut {NoStop}%
\bibitem [{\citenamefont {Scappucci}\ \emph {et~al.}(2020)\citenamefont {Scappucci}, \citenamefont {Kloeffel}, \citenamefont {Zwanenburg}, \citenamefont {Loss}, \citenamefont {Myronov}, \citenamefont {Zhang}, \citenamefont {De~Franceschi}, \citenamefont {Katsaros},\ and\ \citenamefont {Veldhorst}}]{Scappucci2020}%
  \BibitemOpen
  \bibfield  {author} {\bibinfo {author} {\bibfnamefont {G.}~\bibnamefont {Scappucci}}, \bibinfo {author} {\bibfnamefont {C.}~\bibnamefont {Kloeffel}}, \bibinfo {author} {\bibfnamefont {F.~A.}\ \bibnamefont {Zwanenburg}}, \bibinfo {author} {\bibfnamefont {D.}~\bibnamefont {Loss}}, \bibinfo {author} {\bibfnamefont {M.}~\bibnamefont {Myronov}}, \bibinfo {author} {\bibfnamefont {J.-J.}\ \bibnamefont {Zhang}}, \bibinfo {author} {\bibfnamefont {S.}~\bibnamefont {De~Franceschi}}, \bibinfo {author} {\bibfnamefont {G.}~\bibnamefont {Katsaros}},\ and\ \bibinfo {author} {\bibfnamefont {M.}~\bibnamefont {Veldhorst}},\ }\bibfield  {title} {\bibinfo {title} {The germanium quantum information route},\ }\href {https://doi.org/10.1038/s41578-020-00262-z} {\bibfield  {journal} {\bibinfo  {journal} {Nature Reviews Materials}\ }\textbf {\bibinfo {volume} {6}},\ \bibinfo {pages} {926–943} (\bibinfo {year} {2020})}\BibitemShut {NoStop}%
\bibitem [{\citenamefont {Guryanova}\ \emph {et~al.}(2020)\citenamefont {Guryanova}, \citenamefont {Friis},\ and\ \citenamefont {Huber}}]{Guryanova2020}%
  \BibitemOpen
  \bibfield  {author} {\bibinfo {author} {\bibfnamefont {Y.}~\bibnamefont {Guryanova}}, \bibinfo {author} {\bibfnamefont {N.}~\bibnamefont {Friis}},\ and\ \bibinfo {author} {\bibfnamefont {M.}~\bibnamefont {Huber}},\ }\bibfield  {title} {\bibinfo {title} {Ideal projective measurements have infinite resource costs},\ }\href {https://doi.org/10.22331/q-2020-01-13-222} {\bibfield  {journal} {\bibinfo  {journal} {Quantum}\ }\textbf {\bibinfo {volume} {4}},\ \bibinfo {pages} {222} (\bibinfo {year} {2020})}\BibitemShut {NoStop}%
\bibitem [{\citenamefont {Taranto}\ \emph {et~al.}(2023)\citenamefont {Taranto}, \citenamefont {Bakhshinezhad}, \citenamefont {Bluhm}, \citenamefont {Silva}, \citenamefont {Friis}, \citenamefont {Lock}, \citenamefont {Vitagliano}, \citenamefont {Binder}, \citenamefont {Debarba}, \citenamefont {Schwarzhans}, \citenamefont {Clivaz},\ and\ \citenamefont {Huber}}]{Taranto2023}%
  \BibitemOpen
  \bibfield  {author} {\bibinfo {author} {\bibfnamefont {P.}~\bibnamefont {Taranto}}, \bibinfo {author} {\bibfnamefont {F.}~\bibnamefont {Bakhshinezhad}}, \bibinfo {author} {\bibfnamefont {A.}~\bibnamefont {Bluhm}}, \bibinfo {author} {\bibfnamefont {R.}~\bibnamefont {Silva}}, \bibinfo {author} {\bibfnamefont {N.}~\bibnamefont {Friis}}, \bibinfo {author} {\bibfnamefont {M.~P.}\ \bibnamefont {Lock}}, \bibinfo {author} {\bibfnamefont {G.}~\bibnamefont {Vitagliano}}, \bibinfo {author} {\bibfnamefont {F.~C.}\ \bibnamefont {Binder}}, \bibinfo {author} {\bibfnamefont {T.}~\bibnamefont {Debarba}}, \bibinfo {author} {\bibfnamefont {E.}~\bibnamefont {Schwarzhans}}, \bibinfo {author} {\bibfnamefont {F.}~\bibnamefont {Clivaz}},\ and\ \bibinfo {author} {\bibfnamefont {M.}~\bibnamefont {Huber}},\ }\bibfield  {title} {\bibinfo {title} {Landauer versus nernst: What is the true cost of cooling a quantum system?},\ }\href {https://doi.org/10.1103/prxquantum.4.010332} {\bibfield  {journal} {\bibinfo  {journal} {{PRX} Quantum}\
  }\textbf {\bibinfo {volume} {4}} (\bibinfo {year} {2023})}\BibitemShut {NoStop}%
\bibitem [{\citenamefont {Rolandi}\ and\ \citenamefont {Perarnau-Llobet}(2023)}]{rolandi2023finite}%
  \BibitemOpen
  \bibfield  {author} {\bibinfo {author} {\bibfnamefont {A.}~\bibnamefont {Rolandi}}\ and\ \bibinfo {author} {\bibfnamefont {M.}~\bibnamefont {Perarnau-Llobet}},\ }\bibfield  {title} {\bibinfo {title} {Finite-time {L}andauer principle beyond weak coupling},\ }\href {https://doi.org/10.22331/q-2023-11-03-1161} {\bibfield  {journal} {\bibinfo  {journal} {{Quantum}}\ }\textbf {\bibinfo {volume} {7}},\ \bibinfo {pages} {1161} (\bibinfo {year} {2023})}\BibitemShut {NoStop}%
\bibitem [{\citenamefont {Erdman}\ \emph {et~al.}(2023)\citenamefont {Erdman}, \citenamefont {Rolandi}, \citenamefont {Abiuso}, \citenamefont {Perarnau-Llobet},\ and\ \citenamefont {No\'e}}]{Erdman2023Pareto}%
  \BibitemOpen
  \bibfield  {author} {\bibinfo {author} {\bibfnamefont {P.~A.}\ \bibnamefont {Erdman}}, \bibinfo {author} {\bibfnamefont {A.}~\bibnamefont {Rolandi}}, \bibinfo {author} {\bibfnamefont {P.}~\bibnamefont {Abiuso}}, \bibinfo {author} {\bibfnamefont {M.}~\bibnamefont {Perarnau-Llobet}},\ and\ \bibinfo {author} {\bibfnamefont {F.}~\bibnamefont {No\'e}},\ }\bibfield  {title} {\bibinfo {title} {Pareto-optimal cycles for power, efficiency and fluctuations of quantum heat engines using reinforcement learning},\ }\href {https://doi.org/10.1103/PhysRevResearch.5.L022017} {\bibfield  {journal} {\bibinfo  {journal} {Phys. Rev. Res.}\ }\textbf {\bibinfo {volume} {5}},\ \bibinfo {pages} {L022017} (\bibinfo {year} {2023})}\BibitemShut {NoStop}%
\bibitem [{\citenamefont {Rolandi}\ \emph {et~al.}(2023{\natexlab{b}})\citenamefont {Rolandi}, \citenamefont {Abiuso},\ and\ \citenamefont {Perarnau-Llobet}}]{Rolandi2023collective}%
  \BibitemOpen
  \bibfield  {author} {\bibinfo {author} {\bibfnamefont {A.}~\bibnamefont {Rolandi}}, \bibinfo {author} {\bibfnamefont {P.}~\bibnamefont {Abiuso}},\ and\ \bibinfo {author} {\bibfnamefont {M.}~\bibnamefont {Perarnau-Llobet}},\ }\bibfield  {title} {\bibinfo {title} {Collective advantages in finite-time thermodynamics},\ }\href {https://doi.org/10.1103/physrevlett.131.210401} {\bibfield  {journal} {\bibinfo  {journal} {Physical Review Letters}\ }\textbf {\bibinfo {volume} {131}} (\bibinfo {year} {2023}{\natexlab{b}})}\BibitemShut {NoStop}%
\bibitem [{\citenamefont {Abiuso}\ and\ \citenamefont {Perarnau-Llobet}(2020)}]{Abiuso2020}%
  \BibitemOpen
  \bibfield  {author} {\bibinfo {author} {\bibfnamefont {P.}~\bibnamefont {Abiuso}}\ and\ \bibinfo {author} {\bibfnamefont {M.}~\bibnamefont {Perarnau-Llobet}},\ }\bibfield  {title} {\bibinfo {title} {Optimal cycles for low-dissipation heat engines},\ }\href {https://doi.org/10.1103/PhysRevLett.124.110606} {\bibfield  {journal} {\bibinfo  {journal} {Phys. Rev. Lett.}\ }\textbf {\bibinfo {volume} {124}},\ \bibinfo {pages} {110606} (\bibinfo {year} {2020})}\BibitemShut {NoStop}%
\bibitem [{\citenamefont {Weber}(1956)}]{Weber1956}%
  \BibitemOpen
  \bibfield  {author} {\bibinfo {author} {\bibfnamefont {J.}~\bibnamefont {Weber}},\ }\bibfield  {title} {\bibinfo {title} {Fluctuation dissipation theorem},\ }\href {https://doi.org/10.1103/physrev.101.1620} {\bibfield  {journal} {\bibinfo  {journal} {Physical Review}\ }\textbf {\bibinfo {volume} {101}},\ \bibinfo {pages} {1620–1626} (\bibinfo {year} {1956})}\BibitemShut {NoStop}%
\bibitem [{\citenamefont {Kubo}(1966)}]{Kubo1966}%
  \BibitemOpen
  \bibfield  {author} {\bibinfo {author} {\bibfnamefont {R.}~\bibnamefont {Kubo}},\ }\bibfield  {title} {\bibinfo {title} {The fluctuation-dissipation theorem},\ }\href {https://doi.org/10.1088/0034-4885/29/1/306} {\bibfield  {journal} {\bibinfo  {journal} {Reports on Progress in Physics}\ }\textbf {\bibinfo {volume} {29}},\ \bibinfo {pages} {255–284} (\bibinfo {year} {1966})}\BibitemShut {NoStop}%
\bibitem [{\citenamefont {Miller}\ \emph {et~al.}(2019)\citenamefont {Miller}, \citenamefont {Scandi}, \citenamefont {Anders},\ and\ \citenamefont {Perarnau-Llobet}}]{Miller2019}%
  \BibitemOpen
  \bibfield  {author} {\bibinfo {author} {\bibfnamefont {H.~J.~D.}\ \bibnamefont {Miller}}, \bibinfo {author} {\bibfnamefont {M.}~\bibnamefont {Scandi}}, \bibinfo {author} {\bibfnamefont {J.}~\bibnamefont {Anders}},\ and\ \bibinfo {author} {\bibfnamefont {M.}~\bibnamefont {Perarnau-Llobet}},\ }\bibfield  {title} {\bibinfo {title} {Work fluctuations in slow processes: Quantum signatures and optimal control},\ }\href {https://doi.org/10.1103/PhysRevLett.123.230603} {\bibfield  {journal} {\bibinfo  {journal} {Phys. Rev. Lett.}\ }\textbf {\bibinfo {volume} {123}},\ \bibinfo {pages} {230603} (\bibinfo {year} {2019})}\BibitemShut {NoStop}%
\bibitem [{\citenamefont {Denzler}\ \emph {et~al.}(2024)\citenamefont {Denzler}, \citenamefont {Santos}, \citenamefont {Lutz},\ and\ \citenamefont {Serra}}]{Denzler2024}%
  \BibitemOpen
  \bibfield  {author} {\bibinfo {author} {\bibfnamefont {T.}~\bibnamefont {Denzler}}, \bibinfo {author} {\bibfnamefont {J.~F.~G.}\ \bibnamefont {Santos}}, \bibinfo {author} {\bibfnamefont {E.}~\bibnamefont {Lutz}},\ and\ \bibinfo {author} {\bibfnamefont {R.~M.}\ \bibnamefont {Serra}},\ }\bibfield  {title} {\bibinfo {title} {Nonequilibrium fluctuations of a quantum heat engine},\ }\href {https://doi.org/10.1088/2058-9565/ad6287} {\bibfield  {journal} {\bibinfo  {journal} {Quantum Science and Technology}\ }\textbf {\bibinfo {volume} {9}},\ \bibinfo {pages} {045017} (\bibinfo {year} {2024})}\BibitemShut {NoStop}%
\bibitem [{\citenamefont {Denzler}\ and\ \citenamefont {Lutz}(2021)}]{Denzler2021}%
  \BibitemOpen
  \bibfield  {author} {\bibinfo {author} {\bibfnamefont {T.}~\bibnamefont {Denzler}}\ and\ \bibinfo {author} {\bibfnamefont {E.}~\bibnamefont {Lutz}},\ }\bibfield  {title} {\bibinfo {title} {Power fluctuations in a finite-time quantum carnot engine},\ }\href {https://doi.org/10.1103/PhysRevResearch.3.L032041} {\bibfield  {journal} {\bibinfo  {journal} {Phys. Rev. Res.}\ }\textbf {\bibinfo {volume} {3}},\ \bibinfo {pages} {L032041} (\bibinfo {year} {2021})}\BibitemShut {NoStop}%
\bibitem [{\citenamefont {Aggarwal}(2024)}]{kushdata}%
  \BibitemOpen
  \bibfield  {author} {\bibinfo {author} {\bibfnamefont {K.}~\bibnamefont {Aggarwal}},\ }\href {https://doi.org/10.5281/zenodo.14516010} {\bibinfo {title} {Raw data for "rapid optimal work extraction from a quantum-dot information engine"}} (\bibinfo {year} {2024}),\ \bibinfo {note} {[Data set]}\BibitemShut {NoStop}%
\bibitem [{\citenamefont {Jirovec}\ \emph {et~al.}(2021)\citenamefont {Jirovec}, \citenamefont {Hofmann}, \citenamefont {Ballabio}, \citenamefont {Mutter}, \citenamefont {Tavani}, \citenamefont {Botifoll}, \citenamefont {Crippa}, \citenamefont {Kukucka}, \citenamefont {Sagi}, \citenamefont {Martins}, \citenamefont {Saez-Mollejo}, \citenamefont {Prieto}, \citenamefont {Borovkov}, \citenamefont {Arbiol}, \citenamefont {Chrastina}, \citenamefont {Isella},\ and\ \citenamefont {Katsaros}}]{Jirovec2021}%
  \BibitemOpen
  \bibfield  {author} {\bibinfo {author} {\bibfnamefont {D.}~\bibnamefont {Jirovec}}, \bibinfo {author} {\bibfnamefont {A.}~\bibnamefont {Hofmann}}, \bibinfo {author} {\bibfnamefont {A.}~\bibnamefont {Ballabio}}, \bibinfo {author} {\bibfnamefont {P.~M.}\ \bibnamefont {Mutter}}, \bibinfo {author} {\bibfnamefont {G.}~\bibnamefont {Tavani}}, \bibinfo {author} {\bibfnamefont {M.}~\bibnamefont {Botifoll}}, \bibinfo {author} {\bibfnamefont {A.}~\bibnamefont {Crippa}}, \bibinfo {author} {\bibfnamefont {J.}~\bibnamefont {Kukucka}}, \bibinfo {author} {\bibfnamefont {O.}~\bibnamefont {Sagi}}, \bibinfo {author} {\bibfnamefont {F.}~\bibnamefont {Martins}}, \bibinfo {author} {\bibfnamefont {J.}~\bibnamefont {Saez-Mollejo}}, \bibinfo {author} {\bibfnamefont {I.}~\bibnamefont {Prieto}}, \bibinfo {author} {\bibfnamefont {M.}~\bibnamefont {Borovkov}}, \bibinfo {author} {\bibfnamefont {J.}~\bibnamefont {Arbiol}}, \bibinfo {author} {\bibfnamefont {D.}~\bibnamefont {Chrastina}}, \bibinfo {author} {\bibfnamefont {G.}~\bibnamefont
  {Isella}},\ and\ \bibinfo {author} {\bibfnamefont {G.}~\bibnamefont {Katsaros}},\ }\bibfield  {title} {\bibinfo {title} {A singlet-triplet hole spin qubit in planar ge},\ }\href {https://doi.org/10.1038/s41563-021-01022-2} {\bibfield  {journal} {\bibinfo  {journal} {Nature Materials}\ }\textbf {\bibinfo {volume} {20}},\ \bibinfo {pages} {1106–1112} (\bibinfo {year} {2021})}\BibitemShut {NoStop}%
\bibitem [{\citenamefont {Hofmann}\ \emph {et~al.}(2016)\citenamefont {Hofmann}, \citenamefont {Maisi}, \citenamefont {R\"ossler}, \citenamefont {Basset}, \citenamefont {Kr\"ahenmann}, \citenamefont {M\"arki}, \citenamefont {Ihn}, \citenamefont {Ensslin}, \citenamefont {Reichl},\ and\ \citenamefont {Wegscheider}}]{Hofmann2016}%
  \BibitemOpen
  \bibfield  {author} {\bibinfo {author} {\bibfnamefont {A.}~\bibnamefont {Hofmann}}, \bibinfo {author} {\bibfnamefont {V.~F.}\ \bibnamefont {Maisi}}, \bibinfo {author} {\bibfnamefont {C.}~\bibnamefont {R\"ossler}}, \bibinfo {author} {\bibfnamefont {J.}~\bibnamefont {Basset}}, \bibinfo {author} {\bibfnamefont {T.}~\bibnamefont {Kr\"ahenmann}}, \bibinfo {author} {\bibfnamefont {P.}~\bibnamefont {M\"arki}}, \bibinfo {author} {\bibfnamefont {T.}~\bibnamefont {Ihn}}, \bibinfo {author} {\bibfnamefont {K.}~\bibnamefont {Ensslin}}, \bibinfo {author} {\bibfnamefont {C.}~\bibnamefont {Reichl}},\ and\ \bibinfo {author} {\bibfnamefont {W.}~\bibnamefont {Wegscheider}},\ }\bibfield  {title} {\bibinfo {title} {Equilibrium free energy measurement of a confined electron driven out of equilibrium},\ }\href {https://doi.org/10.1103/PhysRevB.93.035425} {\bibfield  {journal} {\bibinfo  {journal} {Phys. Rev. B}\ }\textbf {\bibinfo {volume} {93}},\ \bibinfo {pages} {035425} (\bibinfo {year} {2016})}\BibitemShut {NoStop}%
\end{thebibliography}%
	
	\widetext
	\appendix
    \section{Device fabrication}
    \label{A}
	The devices were processed in the Institute of Science and Technology Austria Nanofabrication facility. A 6x6 mm$^2$ chip comprising of Ge quantum well sandwiched between $\text{Si}_{0.3}\text{Ge}_{0.7}$ is cleaned before further processing. Further details about the quantum well growth can be found in Ref.~\cite{Jirovec2021}. First, ohmic contacts are patterned in a 100 keV electron beam lithography system. Subsequently, a few nanometers of native oxide is milled by argon bombardment which is followed by deposition of 60 nm Pt layer. A 20 nm thick layer of aluminium oxide is deposited at 300$^{\circ}$C in an atomic layer deposition step, followed by gates consisting of 3 nm Ti and 27 nm Pd.
	
	\section{Characterisation of the tunnel rates and lever arm}\label{B}	
	\begin{figure}[h]
		\centering
		\includegraphics{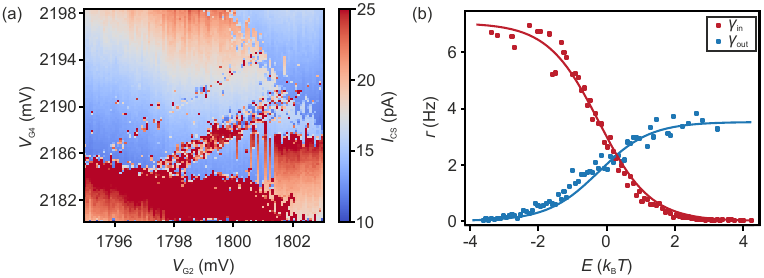}
		\caption{(a) A set of bias triangles measured by varying $V_\text{G2}$ and $V_\text{G4}$ and recording the current $I_\text{CS}$ through the charge sensor at a bias voltage $V_\text{b,QD} = 0.32$ mV. (b) Tunneling rates $\gamma_\text{in}$ and $\gamma_\text{out}$ of QD1.}
		\label{fig:characterisation}
	\end{figure}

    The measurements were performed in a dilution refrigerator at a base temperature of 150 mK.  The charge sensor and the double quantum dot are isolated by creating an electrostatic barrier using splitter gate voltages $V_\text{S1-S2}$. The double quantum dot in the bottom array is tuned by using the gates voltages $V_\text{G2}$ and $V_\text{G4}$. The tunneling between Fermi reservoirs and double quantum dot is controlled using gate voltages $V_\text{G1}$ and $V_\text{G5}$. The inter-dot tunneling is set by the gate voltage $V_\text{G3}$. Similarly, the charge sensor is tuned by a combination of gate voltages $V_\text{CS1-CS3}$. We first measure a set of bias triangles, as shown in \sfref{fig:characterisation}{a}, at a fixed bias $V_\text{b, QD} = 0.32$ mV. We extract a lever arm $\alpha = \frac{ V_\text{b,QD}}{\Delta V_\text{G2}} = 0.041 \pm 0.002$. Time traces across the transition from $n =0$ to $n=1$ were recorded. Tunnel rates and electron temperature were extracted in \sfref{fig:characterisation}{b} using the functions:  $\gamma_\text{in} = \Gamma_\text{in}f(E)$ and $\gamma_\text{out} = \Gamma_\text{out}(1-f(E))$, where $f(E) = (1+e^{E/k_\text{B}T})^{-1}$, which describe the tunneling in and out rates of the quantum dot~\cite{Hofmann2016}. We obtain an electron temperature of $T = 180 $ mK from these fits.  

\section{Calibration drift}\label{C}
\begin{figure}[h]
    \centering
    \includegraphics[width=0.9\textwidth]{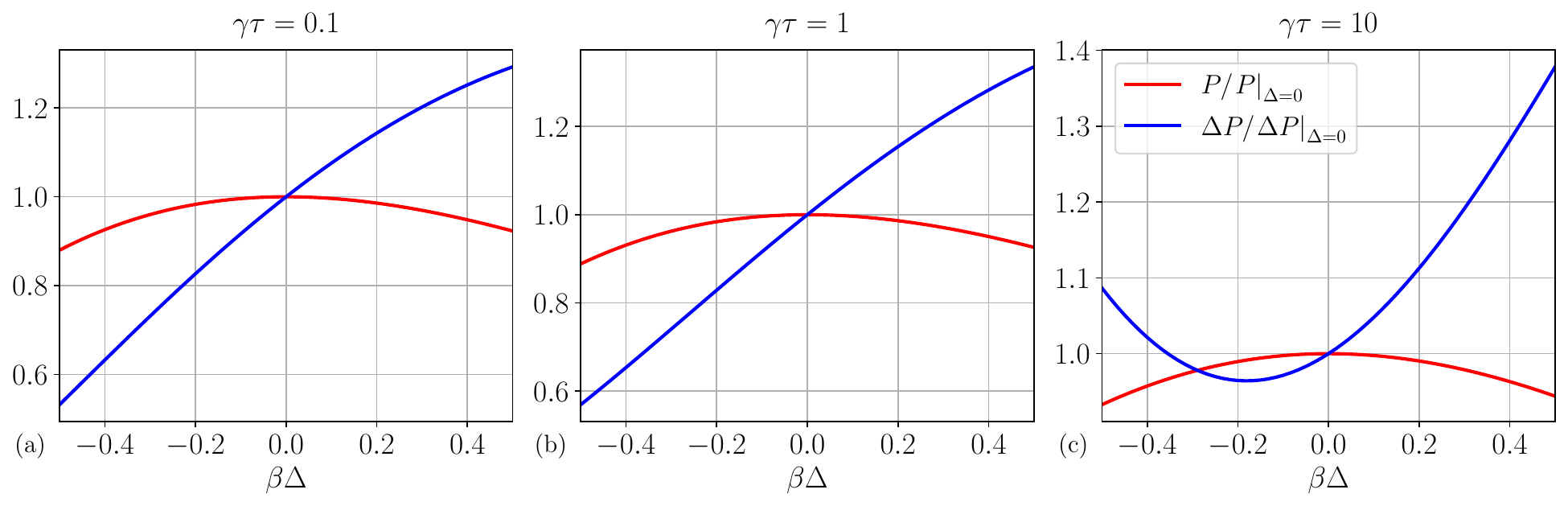}
    \vspace{-10pt}
    \caption{Relative change of power and fluctuations for a constant shift from the optimal protocol $\eps(t) \rightarrow \eps(t) + \Delta$ for all the regimes of driving speeds: (a) $\gamma\tau=0.1$, (b) $\gamma\tau=1$, (c) $\gamma\tau=10$.}
    \label{fig:drift}
\end{figure}
Since the calibration drift takes much longer than the duration of a single cycle (even in the slow driving regime) we can assume that within a cycle it corresponds to a constant shift of the energy $\eps \rightarrow \eps + \Delta$. In \fref{fig:drift} we show how this shift affects the power and fluctuations relative to their value in absence of the shift. As expected, since the power is optimal for $\Delta = 0$, we see that for small shifts it deviates from its optimal value as $\mathcal{O}(\Delta^2)$. Conversely, it is clear that this is not the case for $\Delta P$ since the protocols do not optimise the fluctuations. Therefore its deviations scale as $\mathcal{O}(\Delta)$. This makes the fluctuations more sensitive than the power to this shift caused by the calibration drift.

\section{Measurement statistics}\label{D}
In this section we present how to compute the estimators -- and their variance -- for expected work and work fluctuations that were used for the plotted values and error bars in \fref{fig:main}. 
For $N$ i.i.d statistical samples $\{W_i\}_{i=1}^N$ of a random variable $\mathcal W$. We estimate the expected value of $\mathcal W$ with the mean:
\begin{equation}
    E(\{W_i\}_{i=1}^N) := \frac{1}{N}\sum_{i=1}^N W_i~. 
\end{equation}
It is straightforward to compute the expected value and variance of this estimator:
\begin{align}
    \langle E(\{W_i\}_{i=1}^N)\rangle &= \langle\mathcal W\rangle~,\\
    \mathrm{Var}(E(\{W_i\}_{i=1}^N)) &= \frac{\mathrm{Var}(\mathcal W)}{N}~.
\end{align}
Therefore, with $N$ samples of work gain $\{W_i\}_{i=1}^N$, we compute the expected work with the estimator $E$ and use the square root of its variance for its error in the plots. To compute the variance from the sample, we use the following estimator
\begin{equation}
    V(\{W_i\}_{i=1}^N) = \frac{1}{N-1}\sum_{i=1}^N\left(W_i - E(\{ W_i\}_{i=1}^N)\right)^2~,
\end{equation}
which we can use also to estimate the fluctuations of the work gain. The expected value and variance of this estimator give
\begin{align}
    \langle V(\{W_i\}_{i=1}^N)\rangle &= \mathrm{Var}(\mathcal W)~,\\
    \mathrm{Var}(V(\{W_i\}_{i=1}^N)) &= \langle\mathcal W^4\rangle\frac{N(N-4)+1}{N(N-1)^2} - \frac{4\langle\mathcal W^3\rangle\langle\mathcal W\rangle}{N} - \frac{\mathrm{Var}(\mathcal W)^2}{N-1} + \frac{3\langle\mathcal W^2\rangle^2}{N}~,
\end{align}
where we can use the maximum likelihood estimators for the 2nd, 3rd and 4th moments -- $m_2$, $m_3$ and $m_4$ respectively -- to compute the error of the estimator of the work fluctuations from the statistical samples. These maximum likelihood estimators are given by
\begin{equation}
    m_j(\{W_i\}_{i=1}^N) := \frac{1}{N}\sum_i W_i^j~, 
\end{equation}
and satisfy $\langle m_j(\{W_i\}_{i=1}^N) \rangle = \langle\mathcal W^j\rangle$.
\end{document}